\documentclass[aps, prb, amsmath, twocolumn, longbibliography, superscriptaddress, footinbib, 10pt,floatfix, final]{revtex4-2}

\makeatletter
\def\frontmatter@maketitle{%
  \@author@finish
  \title@column\titleblock@produce
  \suppressfloats[t]%
}%
\makeatother

\newcommand\includeSI{0}

\usepackage{ifthen}
\usepackage{float}
\usepackage{mathtools}
\usepackage{booktabs}
\usepackage[version=4]{mhchem} 
\usepackage{physics}
\usepackage{amssymb, amsmath}
\usepackage{hyphenat}
\usepackage{comment}
\usepackage{graphicx} 
\usepackage{blindtext}
\usepackage{siunitx} 
\usepackage[final]{microtype} 
\usepackage{setspace}
\usepackage[T1]{fontenc}
\usepackage{lmodern}
\usepackage{xcolor}
\usepackage[utf8]{inputenc}
\usepackage[acronym]{glossaries}
\usepackage[l2tabu, orthodox]{nag}
\usepackage{makecell}
\usepackage{color,soul}
\usepackage{adjustbox}
\usepackage{environ} 

\usepackage{hyperref}
\hypersetup{colorlinks,linkcolor=blue, citecolor=blue, urlcolor=blue}

\ifthenelse{\equal{\includeSI}{1}}{
	\input{packages_SI.tex}
}{}

\sethlcolor{yellow}

\graphicspath{{./figures}, {./figures_SI}}

\def\figureautorefname{Fig.} 
\def\tableautorefname{Table} 
\newcommand{\fref}[2]{\autoref{#1}\textcolor{blue}{#2}}

\newcommand{\paperSection}[2][normal]{
    \ifthenelse{\equal{#1}{normal}}{
        \medskip\noindent{\textbf{#2}}\newline
    }{
        \noindent\textbf{#2}\newline
    }
}
\newcommand{\polydept}{Department of Engineering Physics, \'Ecole Polytechnique de Montr\'eal, C.P. 6079, Succ. Centre-Ville, Montr\'eal, Qu\'ebec, Canada H3C 3A7}

\newcommand{\leibnizA}{IHP$-$Leibniz-Institut f\"{u}r innovative Mikroelektronik, Im Technologiepark 25, 15236 Frankfurt (Oder), Germany}

\newcommand{\leibnizB}{Leibniz-Institut f\"{u}r Kristallz\"{u}chtung, Max-Born-Straße 2, 12489 Berlin, Germany}

\newcommand{\GeSn}[2]{Ge$_{#1}$Sn$_{#2}$}
\newcommand{\SiGe}[2]{Si$_{#1}$Ge$_{#2}$}

\DeclareSIUnit{\million}{\text{million}}

\newcommand{\ac}[1]{\gls*{#1}}

\newcommand{\RNum}[1]{\uppercase\expandafter{\romannumeral #1\relax}}

\newcommand\vertarrowbox[3][3ex]{%
  \begin{array}[t]{@{}c@{}} #2 \vspace{1ex}\\
  \left\uparrow\vcenter{\hrule height #1}\right.\kern-\nulldelimiterspace\\
  \makebox[0pt]{#3}
  \end{array}%
}

\NewEnviron{myequation}[1]{%
\begin{equation}
\scalebox{#1}{$\BODY$}
\end{equation}
}

\newcommand{\suppfref}[2]{\autoref{#1}\textcolor{blue}{#2}}

\newcommand\suppFigRef[3]{%
    \ifthenelse{\equal{\includeSI}{1}}{%
        \let\tempfigureautorefname\figureautorefname%
        \renewcommand\figureautorefname{Fig.}%
        \suppfref{#1}{#2}%
        \let\figureautorefname\tempfigureautorefname%
    }{\textcolor{blue}{Fig. S#3}}%
}

\newcommand\suppFigsRef[3]{%
    \ifthenelse{\equal{\includeSI}{1}}{%
        \let\tempfigureautorefname\figureautorefname%
        \renewcommand\figureautorefname{Figs.}%
        \suppfref{#1}{#2}%
        \let\figureautorefname\tempfigureautorefname%
    }{\textcolor{blue}{Figs. S#3}}%
}

\newcommand\suppTabRef[3]{%
    \ifthenelse{\equal{\includeSI}{1}}{%
        \let\temptableautorefname\tableautorefname%
        \renewcommand\tableautorefname{Table}%
        \suppfref{#1}{#2}%
        \let\tableautorefname\temptableautorefname%
    }{\textcolor{blue}{Table S#3}}%
}

\def\blankpage{%
      \clearpage%
      \thispagestyle{empty}%
      \null%
      \clearpage}

\begin{document}


\newacronym{cmos}{CMOS}{complementary metal\hyp{}oxide\hyp{}semiconductor}
\newacronym{eit}{EIT}{electromagnetically\hyp{}induced transparency}
\newacronym{pir}{PIR}{polarization\hyp{}induced resonance}
\newacronym{swir}{SWIR}{short\hyp{}wave infrared}
\newacronym{eswir}{e\hyp{}SWIR}{extended short\hyp{}wave infrared}
\newacronym{aoi}{AOI}{angle of incidence}
\newacronym{ed}{ED}{electric dipole}
\newacronym{md}{MD}{magnetic dipole}
\newacronym{mir}{MIR}{mid infrared}
\newacronym{nwa}{NWA}{nanowire array}
\newacronym{nw}{NW}{nanowire}
\newacronym{fdtd}{FDTD}{finite\hyp{}difference time domain}
\newacronym{fom}{FoM}{figure\hyp{}of\hyp{}merit}
\newacronym{pbc}{PBC}{periodic boundary conditions}
\newacronym{pml2}{PML}{perfectly matched layers}
\newacronym{sfs}{SFs}{stacking faults}
\newacronym{cho}{CHO}{coupled harmonic oscillator}
\newacronym{ri}{RI}{refractive index}
\newacronym{fwhm}{FWHM}{full\hyp{}width half\hyp{}maximum}

\newacronym{eels}{EELS}{electron energy loss spectroscopy}
\newacronym{tem}{TEM}{transmission electron microscopy}
\newacronym{sem}{SEM}{scanning electron microscope}
\newacronym{stem}{STEM}{scanning transmission electron microscopy}
\newacronym{rsm}{RSM}{reciprocal space mapping}
\newacronym{xrd}{XRD}{X\hyp{}ray diffraction}
\newacronym{hrtem}{HRTEM}{High\hyp{}resolution transmission electron microscopy}
\newacronym{ebl}{EBL}{electron\hyp{}beam lithography}
\newacronym{rie}{RIE}{reactive\hyp{}ion etching}
\newacronym{cvd}{CVD}{low\hyp{}pressure chemical vapor deposition}

\newacronym{acf}{ACF}{autocorrelation function}
\newacronym{afm}{AFM}{atomic force microscopy}
\newacronym{ft}{FT}{Fourier transform}
\newacronym{csnw}{CSNW}{core/shell NW}
\newacronym{tfsf}{TFSF}{total\hyp{}field/scattered\hyp{}field}
\newacronym{pml}{PML}{Perfectly matched layer}
\newacronym{fft}{FFT}{Fast\hyp{}Fourier transform}

\title{Polarization\hyp{}Tuned Fano Resonances in All\hyp{}Dielectric Short\hyp{}Wave Infrared Metasurface}

\author{Anis Attiaoui}
\affiliation{\polydept{}}

\author{G{\'{e}}rard Daligou}
\affiliation{\polydept{}}

\author{Simone Assali}
\affiliation{\polydept{}}

\author{Oliver Skibitzki}
\affiliation{\leibnizA{}}

\author{Thomas Schroeder}
\affiliation{\leibnizB{}}

\author{Oussama Moutanabbir}
\affiliation{\polydept{}}

\begin{abstract}
\medskip
The \ac{swir} is an underexploited portion of the electromagnetic spectrum in metasurface\hyp{}based nanophotonics despite its strategic importance in sensing and imaging applications. This is mainly attributed to the lack of material systems to tailor light\hyp{}matter interactions in this range. Herein, we address this limitation and demonstrate an all\hyp{}dielectric silicon\hyp{}integrated metasurface enabling polarization\hyp{}induced Fano resonance control at \ac{swir} frequencies. The platform consists of a two\hyp{}dimensional \ce{Si}/\GeSn{0.9}{0.1} core/shell nanowire array on a silicon wafer. By tuning the light polarization, we show that the metasurface reflectance can be efficiently engineered due to Fano resonances emerging from the electric and magnetic dipoles competition. The interference of optically induced dipoles in high\hyp{}index nanowire arrays offers additional degrees of freedom to tailor the directional scattering and the flow of light while enabling sharp polarization\hyp{}modulated resonances. This tunability is harnessed in nanosensors yielding an efficient detection of $10^{-2}$ changes in the refractive index of the surrounding medium.\\

\textbf{Keywords:} \ac{swir}, Metasurface, Polarization tunability, Fano resonance, Si/\GeSn{}{} core/shell, Nanowire,  Refractive index sensing \\
\end{abstract}

\maketitle

All\hyp{}dielectric nanostructures are versatile platforms to engineer light\hyp{}matter interactions and exploit a range of processes including Fano interference \cite{Fan2014OpticalNanostructure,Limonov2017FanoPhotonics} and strong coupling \cite{Sarma2021}. Their fundamental properties support both \ac{ed} and strong \ac{md} resonances in addition to higher\hyp{}order multipole resonances, which substitute lossy ohmic currents with low\hyp{}loss displacement currents emerging from bounded electrons oscillations \cite{Ginn2012RealizingMetamaterials,Moitra2013RealizationMetamaterial,Fan2014OpticalNanostructure}. This is in contrast to plasmonic particles, where a non\hyp{}negligible magnetic response can only be achieved using complex geometries \cite{Linden2004MagneticTerahertz,Enkrich2005}. Nonetheless, the widely used materials for metasurfaces still remain radiating metallic antennas \cite{Soukoulis2011PastMetamaterials} despite numerous challenges \cite{Zhou2005SaturationFrequencies}. For instance, the significant non\hyp{}radiative conductor losses usually lead to a broad bandwidth $(>\SI{50}{\nano\meter})$ and restrict the achievable $Q$\hyp{}factor to less than $\sim 10$ \cite{He2019AnalogueCoupling}. Additionally, the anisotropic electromagnetic response impedes the control of the resonance. In fact, given that metasurface geometrical design influences the spatial overlap of resonance\hyp{}related field distributions, it becomes challenging to dynamically modulate the resonant features with a fixed design structure.\\
\indent It is thus highly coveted to develop low\hyp{}loss resonators made entirely of dielectrics with average permittivity in the $\approx 2-14$ range, whilst still being sufficiently subwavelength in the propagation direction. In this regard, a plethora of materials has been explored \cite{hao2008, zhang2013, Zhu2020Realization5,Zhang2021ElectricallyMaterial} enabling novel photonic functionalities \cite{meng2021,chen2016,staude2017, Krasnok2017}. The resonant multipole interplay in all\hyp{}dielectric metasurfaces can induce sharp features in light reflection and transmission spectra, including Fano resonances \cite{Limonov2017FanoPhotonics}. The characteristic asymmetric Fano line shape is due to the coherent interference between two hybridized broad and narrow scattering pathways \cite{Limonov2017FanoPhotonics,Lukyanchuk2010TheMetamaterials}. Tuning these resonances mainly rely on changing the asymmetry parameter, the geometry, or the coupling distance within the unit cell of the metasurface \cite{Cao2012Low-lossMetamaterials,Offermans2011UniversalSensors}. Indeed, current approaches to modulate Fano resonances generally work in passive mode \cite{Yang2015}. Therefore, it is critical to establish alternative design routes for active control of metasurfaces. One possibility is to change the incident light polarization to alter the resonance loss \cite{Hu2020}. Thus, a dynamic control of resonances can be achieved by rotating the corresponding field nodes of the respective modes to ultimately tune their coupling efficiency.\\ 
\indent Herein, we propose a new material system to simultaneously address the aforementioned challenges. This platform leverages the flexibility offered by the emerging silicon-compatible \GeSn{}{} semiconductors \cite{Moutanabbir2021} to implement all\hyp{}dielectric core/shell \ce{Si}/\GeSn{}{} base\hyp{}tapered \ac{nw}\hyp{}metasurfaces, where the polarization-modulated response is controlled in the heretofore unexplored \ac{swir} range. By rotating the polarization state of incident light, \ac{pir} modes are observed and finely controlled with a modulation depth as high as 75\%. Furthermore, by combining the narrow Fano resonance linewidth with the strong near\hyp{}field confinement, we demonstrate a room\hyp{}temperature optical \ac{ri} nanosensor operating between 1.6 and $\SI{1.9}{\um}$, with a sensitivity as high as $\SI{386}{\nm}$/RIU and a \ac{fom} of 12. The Fano resonance spectral shift is induced by a $10^{-2}$ change in the properties of the surrounding environment \ac{ri}.

\onecolumngrid
\adjustimage{width=.85\textwidth,center,caption={\textbf{Configuration of the \ce{Si}/\GeSn{}{} NW metasurface allowing polarization-enabled modulation of Fano resonance.} (a) $3D$ schematic of the proposed core/shell array metasurface where the dynamic modulation of the polarization state of the incident light is demonstrated. The polarization and incidence angle, $\varphi$ and $\theta$ , are also shown. (b) The relevant geometrical parameters of the FDTD simulation domain of the tapered NW are indicated: the height \textit{H}, top shell diameter $\textit{d}_{sT}$, bottom shell diameter $\textit{d}_{sB}$, top core diameter $\textit{d}_{cT}$, bottom core diameter $\textit{d}_{cB}$, overgrown \GeSn{}{} layer $\textit{t}$, and top base $\textit{a}$, bottom base $\textit{b}$ and height of the trapezoid $\textit{H}$. The geometrical parameters are presented in the \textcolor{blue}{Table S4}. (c) Measured and FDTD simulated specular reflectance spectra of the Si/\GeSn{}{} core/shell at two different polarizations (\textit{p}$\equiv(\varphi=\SI{0}{\degree})$ and \textit{s}$\equiv(\varphi=\SI{90}{\degree})$). The black arrow indicates the peak position where the reflectance is suppressed. SEM images (tilting angle of $\SI{45}{\degree}$) of (d) the Si core NW template, where the inset indicated the tapered NW geometry. The Si/\GeSn{}{} core/shell NW images are also shown in panel (e) where the zoom in map highlights the tapered base. The associated individual EELS maps for \ce{Si}, \ce{Ge}, and \ce{Sn} atoms are also also shown. The scale bar in all images is fixed to $\SI{200}{\nm}$.},label={fig:fig1meta},nofloat=figure,vspace=\bigskipamount}{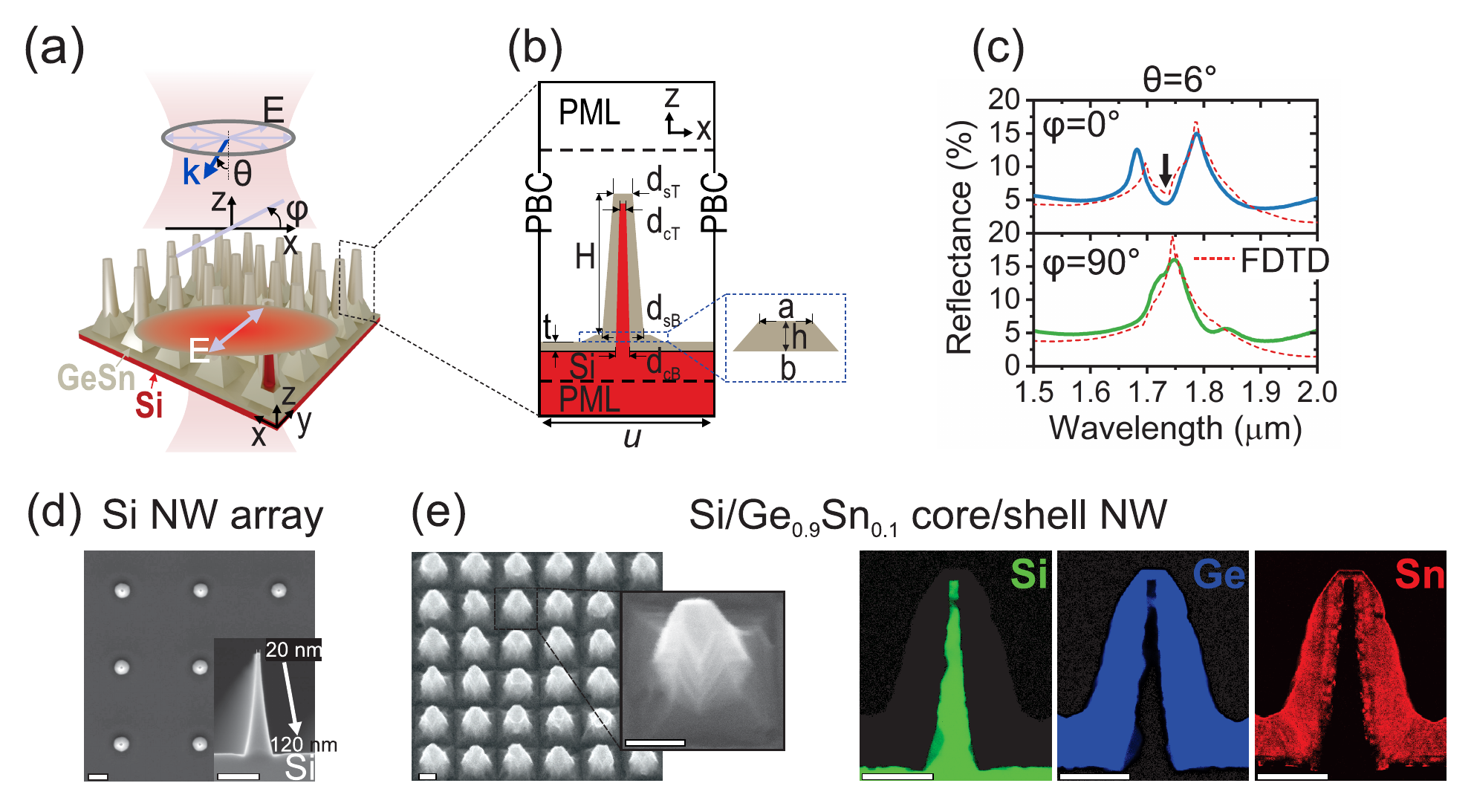}

\twocolumngrid

\UseRawInputEncoding
\par \textbf{Design and characterization.} 
The schematic of the designed semiconductor metasurface is shown in \fref{fig:fig1meta}{a}. The unit cell consists of a core/shell \ce{Si}/\GeSn{0.9}{0.1} tapered \ac{nw}. The \ac{nw} dimensions are chosen such that the resonance of both \ac{ed} and \ac{md} modes are closely aligned in frequency. The geometrical configuration of a single unit cell is illustrated in \fref{fig:fig1meta}{b}. Numerical simulations of the fabricated structures are carried out using a $3D$ \ac{fdtd} solver (\textcolor{blue}{Methods}). The corresponding simulated and measured reflectance spectra are displayed in \fref{fig:fig1meta}{c} for two different incident light polarization (\textit{p}\hyp{}type: $\varphi=\SI{0}{\degree}$, and \textit{s}\hyp{}type: $\varphi=\SI{90}{\degree}$). A distinct \ac{pir}\hyp{}like dip, accompanied with two peaks, is observed at $\SI{1734}{\nm}$ for $\varphi=\SI{0}{\degree}$. When the electric field is polarized along the z\hyp{}axis, $\varphi=\SI{90}{\degree}$, a narrow reflectance peak with a \ac{fwhm} of $\SI{32}{\nm}$ appears. The reflectance amplitude decreases from 17\% to 5\% as $\varphi$ decreases from $\SI{90}{\degree}$ to $\SI{0}{\degree}$. The measured reflectance is acquired by illuminating the sample with a near\hyp{}normal incident $(\SI{6}{\degree})$ broadband white light with a tunable incident light polarization (\textcolor{blue}{Methods}). Note that the measured specular reflectance of the core/shell array is lower than that of bare \ce{Si} \ac{nw} array in the wavelength range between $\SI{1.1}{\um}$ and $\SI{2.5}{\um}$ (\textcolor{blue}{Fig. S1}). A \GeSn{0.9}{0.1} thin film has an estimated electronic band gap, of $\sim \SI{0.65}{\electronvolt}$ $(\approx \SI{1.9}{\um})$ \cite{Attiaoui2021}. The low reflection can be attributed to light absorption within the array. The reflectance $\textit{R}$ is low over the whole spectral and angular range, with a maximum reflectance above 13\% for $\lambda \geqslant \SI{2.25}{\um}$. 
\begin{figure*}[htp]
    \centering
    \includegraphics[width=.55\textwidth,keepaspectratio]{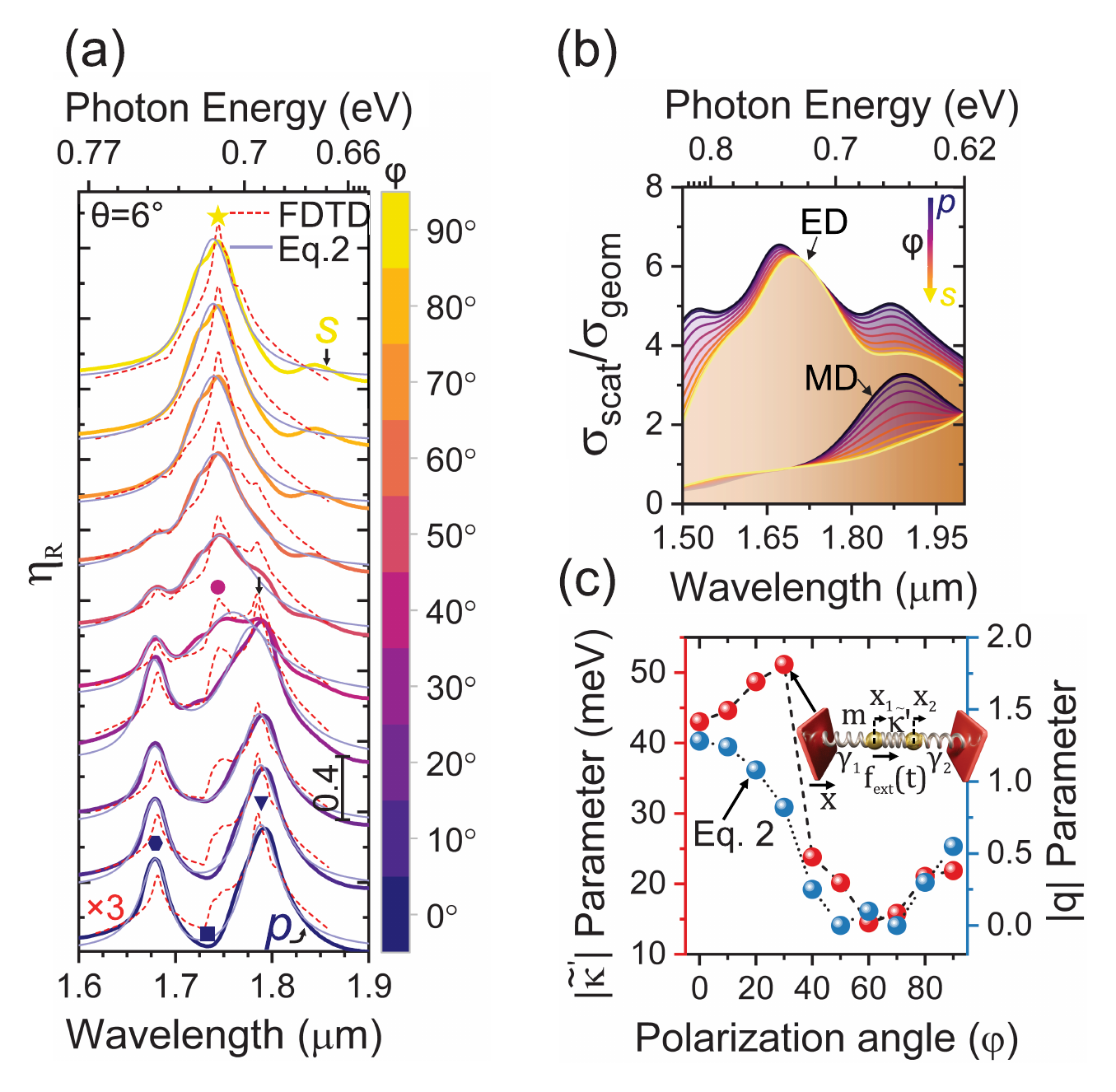}
    \caption{\textbf{Light polarization and Fano resonance.} (a) Polarization\hyp{}dependent reflectance ratio $\eta_{\textit{R}}$ characterization at a fixed \ac{aoi} of $\SI{6}{\degree}$. The polarization is rotated from \textit{s}\hyp{}to \textit{p}\hyp{}state with a $\SI{10}{\degree}$ step. The simulated \ac{fdtd} specular reflectance ratio at each polarization is also overlaid in red\hyp{}dashed traces. The theoretical model based on the lineshape fit (Eq. \ref{eqn:Rfit}) is also shown as gray trace, whereas the CHO fit is shown in the \textcolor{blue}{Fig. S7}. The different spectra are shifted upwards with 0.4 for clarity. The different markers are a visual reference for the near-field distribution that are analyzed in \fref{fig:fig3theory}{}. (b) Scattering cross\hyp{}section of the core/shell \ce{Si}/\GeSn{0.9}{0.1} \ac{nw} as a function of the polarization. The \ac{ed} and \ac{md} modes evolution vs. polarization angle $\varphi$ is highlighted. (c) The evolution of the coupling parameter $\abs{\tilde{\kappa}'}$ and the asymmetry parameter $(\abs{q})$ as a function of $\varphi$ evaluated respectively based on Eqs. \ref{eqn:Rfit} and \ref{eqn:CHO}. The inset shows a schematic of the mechanical coupled harmonic oscillator model used to simulate the \ac{pir} based on Eq. \ref{eqn:CHO}. Only the left oscillator is excited by the external force.}
    \label{fig:fig2eit}
\end{figure*}
\par The proposed nanostructure (\fref{fig:fig1meta}{b}) is composed of: \ce{Si} substrate, $\sim \SI{100}{\nm}$\hyp{}thick \GeSn{}{} layer, \GeSn{0.9}{0.1} trapezoidal base, tapered \ce{Si} core, and \GeSn{0.9}{0.1} shell. The length and the pitch $(u)$ of the \ce{Si} core are $\sim \SI{485}{\nm}$ and $\SI{500}{\nm}$, respectively (\fref{fig:fig1meta}{d}). The core diameter increases toward the base from $\SI{20}{\nm}$ to $\SI{120}{\nm}$ (tapering angle of $\SI{6}{\degree}$). The refractive index of \ce{Si} and \GeSn{0.9}{0.1} thin film are measured with spectroscopic ellipsometry \cite{Attiaoui2021}. Next, the designed structure was built in two separate steps (\textcolor{blue}{Methods}). The pitch was varied from $\SI{500}{\nm}$ to $\SI{2}{\um}$. A \ac{sem} image of the metasurface and the corresponding \ac{eels} images of the \ce{Si}, \ce{Ge}, and \ce{Sn} elements acquired using \ac{tem} are displayed in \fref{fig:fig1meta}{e}. 
\par Rotating the incident light polarization from \textit{p}\hyp{} to \textit{s}\hyp{} state induces a clear reflectance enhancement from 5\% to 17\% at $\SI{1.734}{\um}$, as shown in \fref{fig:fig1meta}{c} (indicated by the black arrow). This is equivalent to a transmission reduction from 50\% to less than 5\%, as simulated with \ac{fdtd} (\textcolor{blue}{Fig. S4}). The \ce{Si} core arrays were fabricated on single\hyp{}side polished wafers, which prevents transmission measurements. Thus, reflectance $(\textit{R})$ measurements are the focus of this study.  The absorption measurements confirmed that the observed spectral features are emanating from the metasurface (\textcolor{blue}{Fig. S5}). To study the optical reflectance of the core/shell \ac{nw}s independently from the optical effects of the \ce{Si} core, the ratio of the core/shell to that of the \ce{Si} arrays $\eta_{\textit{R}}= R_{\text{core/shell\;NW}}/R_{\text{Si\;NW}}$ is analyzed. Noticeably, the lineshape asymmetry is greatly affected by the polarization state of the incident light. The general lineshape of Fano\hyp{}like resonances describes a broad variety of nanostructures \cite{Gallinet2011AbMetamaterials}. To that end, polarization\hyp{}dependent specular reflectance is analyzed to elucidate the physical mechanism of the \ac{pir} features, and the lineshape evolution. \fref{fig:fig2eit}{a} details the ratio $\eta_{\textit{R}}$ at a fixed \ac{aoi} of $\SI{6}{\degree}$ where the electric field orientation is rotated from \textit{s} to \textit{p} polarization with a $\SI{10}{\degree}$ step. As $\varphi$ decreases from $\SI{90}{\degree}$ to $\SI{0}{\degree}$, the reflectance peak at $\SI{1.74}{\um}$ is completely suppressed. Theoretical modeling is introduced to describe quantitatively the \ac{pir} response. 
\par \textbf{Theoretical Model.} Polarization\hyp{}dependent scattering cross\hyp{}section simulations are undertaken for a single \ce{Si}/\GeSn{0.9}{0.1} \ac{nw} to reveal how light\hyp{}matter interactions drive the \ac{pir} effect. A multipole expansion approach is performed using the theoretical method developed by Alaee \textit{et al.} \cite{Alaee2018AnApproximation}, where the impact of polarization on the multipole moments and their contributions to the total scattering cross\hyp{}sections can be estimated. These multipole moments, computed using the electric field and the refractive index extracted from the $3\text{D}$ \ac{fdtd} simulations, are used to evaluate the corresponding scattering cross\hyp{}sections. \fref{fig:fig2eit}{b} shows the evolution of the \ac{ed} and \ac{md} moments as $\varphi$ is gradually changed from \textit{p} to \textit{s}\hyp{}state. The magnetic and electric quadruple amplitudes are 13 and 4\hyp{}fold smaller than their \ac{md} and \ac{ed} counterparts, respectively (\textcolor{blue}{Fig. S6}). Hence, the interference between the optically induced \ac{md} and \ac{ed} is the driving force behind the observed \ac{pir} features. Additionally, as $\varphi$ increases from $\SI{0}{\degree}$ to $\SI{90}{\degree}$, the \ac{md} mode near $\SI{1.890}{\um}$ is completely suppressed, whereas the \ac{ed} mode at $\SI{1.670}{\um}$ redshifts $\SI{25}{\nm}$, and a small mode near $\SI{1.890}{\um}$ still remains. It is important to highlight that the modal response has a Fano\hyp{}like lineshape. This is confirmed by fitting the reflectance ratio $\eta_R$ to a Fano\hyp{}like asymmetric lineshape \cite{Gallinet2011AbMetamaterials} (\textcolor{blue}{Figs. S7-S8}).
\par A closed\hyp{}form analytical formula \cite{Gallinet2011AbMetamaterials} is used to extract the relevant \ac{pir} Fano parameters based on lineshape fitting (Eq. \ref{eqn:Rfit}, \textcolor{blue}{Methods}). Noticeably, Eq. \ref{eqn:Rfit} assumes $\omega_2\gg\gamma_2$. Therefore, the Fano resonance results from the competition between the two dipoles, and the Fano lineshape heavily depends on the coupling between the relative dipole strengths. Consequently, the quantity $q$ is a measure of this relative strength and, based on Fano's original theory \cite{Fano1961}, it can be expressed as:  
\begin{myequation}{1.1}
    \begin{aligned}
    q = \frac{1}{\pi L g}\times\frac{\mu_{\text{MD}}}{\mu_{\text{ED}}}
    \label{eqn:asymparam}
    \end{aligned}
\end{myequation}
where $L$ is the electromagnetic density of states at the \ac{ed} resonance, $g$ is the coupling strength between the \ac{ed} and \ac{md} modes, and $\mu_{\text{ED}}$ and $\mu_{\text{MD}}$ are the total electric and magnetic moments. From \fref{fig:fig2eit}{b}, the \ac{ed} has a stronger moment than that of \ac{md} below $\SI{1.7}{\um}$, which causes $\abs{q}\approx1$ and the asymmetrical Fano lineshapes become distinguishable at low polarization angle $(\varphi<\SI{20}{\degree})$. The experimental and simulated data indicate that $q$ is negative in our system (\textcolor{blue}{Table S1}). The fitting curves are shown in \fref{fig:fig2eit}{a} as gray traces. As $\varphi$ increases from $\SI{30}{\degree}$ to $\SI{90}{\degree}$, $\abs{q}$ reaches a minimum value of $0$ at $\SI{50}{\degree}$, and the asymmetrical lineshape becomes indistinguishable. The evolution of $\abs{q}$, extracted from Eq. \ref{eqn:Rfit}, as a function of the polarization angle $(\varphi)$ is shown in \fref{fig:fig2eit}{c} (blue circles). From Eq. \ref{eqn:asymparam}, the Fano lineshape also depends on $L$ and $\mu_{\text{ED}}$ of the \ac{ed} resonance, and the coupling strength $g$. Therefore, changes to the metasurface geometry can tune the Fano lineshape asymmetry as well. To reveal how this works, we simulated the \textit{p}\hyp{} polarized specular reflectance at $\SI{6}{\degree}$ \ac{aoi} of different metasurfaces with variables bottom shell diameter $d_{sB}$ (from $\SI{150}{\nm}$ to $\SI{350}{\nm}$) and a fixed top shell diameter $d_{sT}$ of $\SI{80}{\nm}$. The change in the lineshape asymmetry is evident as $d_{sB}$ increases (\textcolor{blue}{Fig. S9}). In fact, $q$ increases from $-1.76$ to $0.30$, confirming the geometrical Fano resonance tunability.
\par To analyze the coupling strength $g$ in the \ce{Si}/\GeSn{}{} metasurface, which is inaccessible with Fano's original formalism, we used the \ac{cho} model to simulate the reflectance spectra \cite{GarridoAlzar2002ClassicalTransparency,Lukyanchuk2010TheMetamaterials,Riffe2011ClassicalOscillator}. The equivalent system shown in the inset of \fref{fig:fig2eit}{c} is described by two coupled second\hyp{}order differential equations, subject to an external source $f_{ext}(t)$ as indicated in Eq. \ref{eqn:CHO} (\textcolor{blue}{Methods}). The \ac{cho} models exceptionally well the Fano resonance (\textcolor{blue}{Fig. S7}). The coupling coefficient $\tilde{\kappa}'$ parameter is, in principle, related to the coupling strength $g$ in Eq. \ref{eqn:asymparam}. Based on \fref{fig:fig2eit}{c} (red circles), the \ac{ed} and \ac{md} coupling parameter $(\tilde{\kappa}'\sim g)$ show a clear drop from $\SI{50}{\meV}$ to $\SI{25}{\meV}$ at $\varphi=\SI{40}{\degree}$, indicating a reduction in the transition dipole moments. Furthermore, the same qualitative trend is observed with the asymmetry parameter $(\abs{q})$, where $\abs{q}$ decreases from 1.28 to 0.25 to reach 0 at $\varphi=\SI{50}{\degree}$. This confirms the direct relationship between the Fano resonance lineshape and the coupling strength, as stated in Eq. \ref{eqn:asymparam}.
\begin{figure*}[htp]
    \centering
    \includegraphics[width=.8\textwidth]{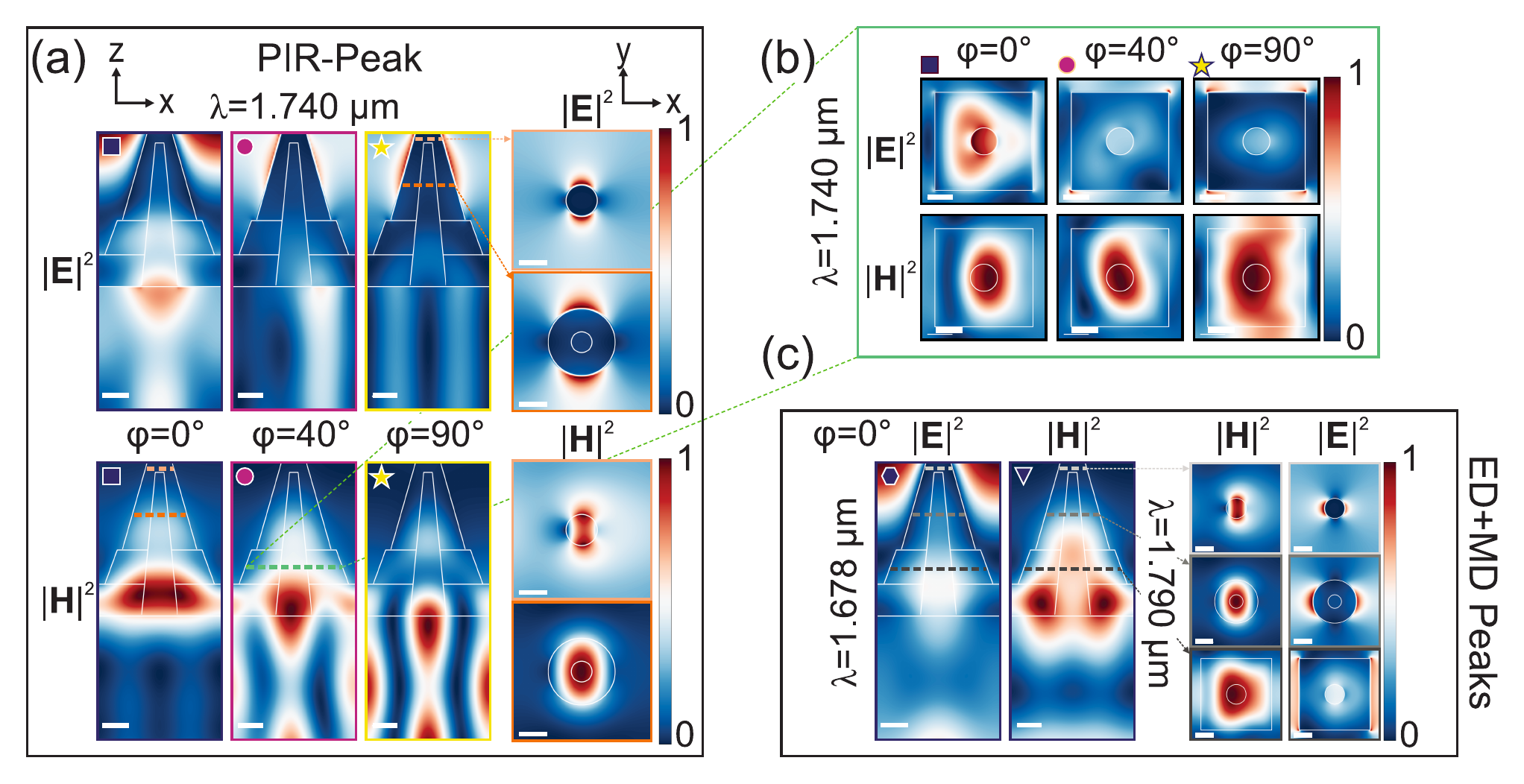}
    \caption{\textbf{Effect of polarization on the near\hyp{}field electric and magnetic distributions.} \ac{fdtd} simulated electric and magnetic field distribution maps in the $y=0$ and $(x\hyp{}y)$ planes for the \ce{Si}/\GeSn{}{} array at specific spectral resonances: (a-b) at the polarization\hyp{}induced reflectance dip located at $\SI{1.740}{\um}$ and (c) the \ac{ed} and \ac{md} resonances, near $\SI{1.678}{\um}$ and $\SI{1.790}{\um}$, respectively. In (b), the effect of incident light polarization is investigated through a systematic description of the electric and magnetic field near only the base of the \ac{nw}. The scale bar is set to $\SI{100}{\nm}$ in all maps.}
    \label{fig:fig3theory}
\end{figure*}
\par To evaluate near\hyp{}field coupling around the \ac{pir}, the near\hyp{}field distributions $(\abs{\mathbf{E}}^2, \abs{\mathbf{H}}^2)$ of the array are systematically analyzed at different polarization. Both electric and magnetic moments are highly polarization\hyp{}dependant ($\cos{\varphi}$ and $\sin{\varphi}$ dependence respectively). Therefore, at the \ac{pir}, the \ac{ed} and \ac{md} can be selectively enhanced. \fref{fig:fig3theory}{} shows the normalized near\hyp{}field distributions at two different planes: $y=0$ and $(x\hyp{}y)$ for a \ac{nw} unit cell at three different polarizations. \fref{fig:fig3theory}{a,b} show the near\hyp{}field distributions at the \ac{pir} mode, at $\SI{1.740}{\um}$, whereas \fref{fig:fig3theory}{c} at the (\ac{ed}, \ac{md}) modes, at $\SI{1.678}{\um}$ and $\SI{1.790}{\um}$ respectively. First, the $(x\hyp{}y)$ plane field distribution maps are shown at specific $z$ values of the \ac{nw} height. The heights are color\hyp{}coded with the dashed traces in panel (a) for $\abs{\mathbf{E}}^2$ at $\varphi=\SI{90}{\degree}$ and for $\abs{\mathbf{H}}^2$ at $\varphi=\SI{0}{\degree}$. Second, a strong magnetic field localization at $\SI{1.740}{\um}$ is observed at the base of the \ac{nw} when $\varphi$ decreases from $\SI{90}{\degree}$ to $\SI{0}{\degree}$. Third, in \fref{fig:fig3theory}{b}, the normalized near\hyp{}field distribution of the \ac{nw} is analyzed near the base of the \ac{nw} (\fref{fig:fig3theory}{b}), at $\SI{1.740}{\um}$ for three polarization angles $\varphi=\SI{0}{\degree},\SI{40}{\degree}$ and $\SI{90}{\degree}$. \ac{ed} and \ac{md} are dominant at $\varphi=\SI{0}{\degree}$, which is corroborated with the scattering cross\hyp{}section calculations shown in \fref{fig:fig2eit}{b}. When $\varphi=\SI{90}{\degree}$, the electric field distribution shows an \ac{ed} behavior with an enhanced field near the edges of the base, whereas the magnetic field $\abs{\mathbf{H}}^2$ loses its \ac{md} character which is confirmed through the simulated large modal broadening near the base. Next, when $\varphi=\SI{40}{\degree}$, \ac{md} and \ac{ed} are equally excited and the near\hyp{}field shows hybrid features. The nature of the electric and magnetic dipoles is well demonstrated in the $(x\hyp{}y)$ cross\hyp{}section maps of $\abs{\mathbf{E}}^2$ at $\SI{1.678}{\um}$ and $\abs{\mathbf{H}}^2$ at $\SI{1.790}{\um}$ (\fref{fig:fig3theory}{c}), as the near\hyp{}field shows typical dipole distribution in the top and middle region of the \ac{nw} \cite{Bohren1998AbsorptionParticles}. Near the base of the \ac{nw}, the electric field is enhanced inside the \ce{Si} core and the \GeSn{}{} base edge, whereas the magnetic field is mainly confined inside the base of the \ac{nw}.
According to \fref{fig:fig3theory}{}, the near\hyp{}field distributions can be effectively manipulated by tuning the polarization of the incident light. The electric field can be selectively enhanced (\fref{fig:fig3theory}{b}) at $\varphi=\SI{0}{\degree}$ or suppressed at $\varphi=\SI{90}{\degree}$ near the base of the \ac{nw}. Thus, the electric and magnetic dipole interference remains polarization sensitive, which promotes the Fano resonance. For different polarizations, the \ac{ed} and \ac{md} can also interfere simultaneously with the geometric resonance of the array through diffractive coupling thereby enhancing polarization\hyp{}dependent Fano resonances.
\par To further inspect diffraction coupling effect on the \ac{pir} modes, period\hyp{}dependent reflectance ratio of the core/shell metasurface is investigated. Three additional \ce{Si}/\GeSn{}{} arrays are fabricated with variable periods of $\SI{800}{\nm}$, $\SI{1}{\um}$, and $\SI{2}{\um}$ as shown in the \ac{sem} images of \textcolor{blue}{Fig. S10}. Given that only four periods are built, complementary \ac{fdtd} simulations are undertaken for an increasing pitch length from $\SI{400}{\nm}$ to $\SI{2}{\um}$. Increasing the period induces a redshift and spectral broadening of the \ac{ed} and \ac{md} modes up to $\SI{575}{\nm}$. This indicates that diffraction coupling to the dipole moments plays an important role in defining the width of the Fano resonance. The combined effect of diffraction coupling and the Fano resonance asymmetry shapes the nature of the observed \ac{pir}. This finding lays the groundwork to exploit the \ac{nw} metasurface in sensing applications, as demonstrated below.
\par \textbf{Refractive index sensing.} Due to the sharp Fano resonance between $\SI{1.60}{\um}$ and $\SI{1.9}{\um}$ and its polarization\hyp{}enabled tunability, one promising application of these metasurfaces is optical sensing. The performance of \ac{ri} sensing is typically evaluated by the \ac{fom}, defined as the ratio of the sensitivity $(S)$ to the \ac{fwhm} at the resonant peak \cite{Sherry2005LocalizedNanocubes},
\begin{myequation}{1.3}
    \text{FoM} = \frac{S\;(nm/\mathrm{RIU})}{\mathrm{\ac{fwhm}}\;(nm)}
    \label{eqn:fom}
\end{myequation}
\begin{figure*}[htb]
    \centering
    \includegraphics[width=.9\textwidth]{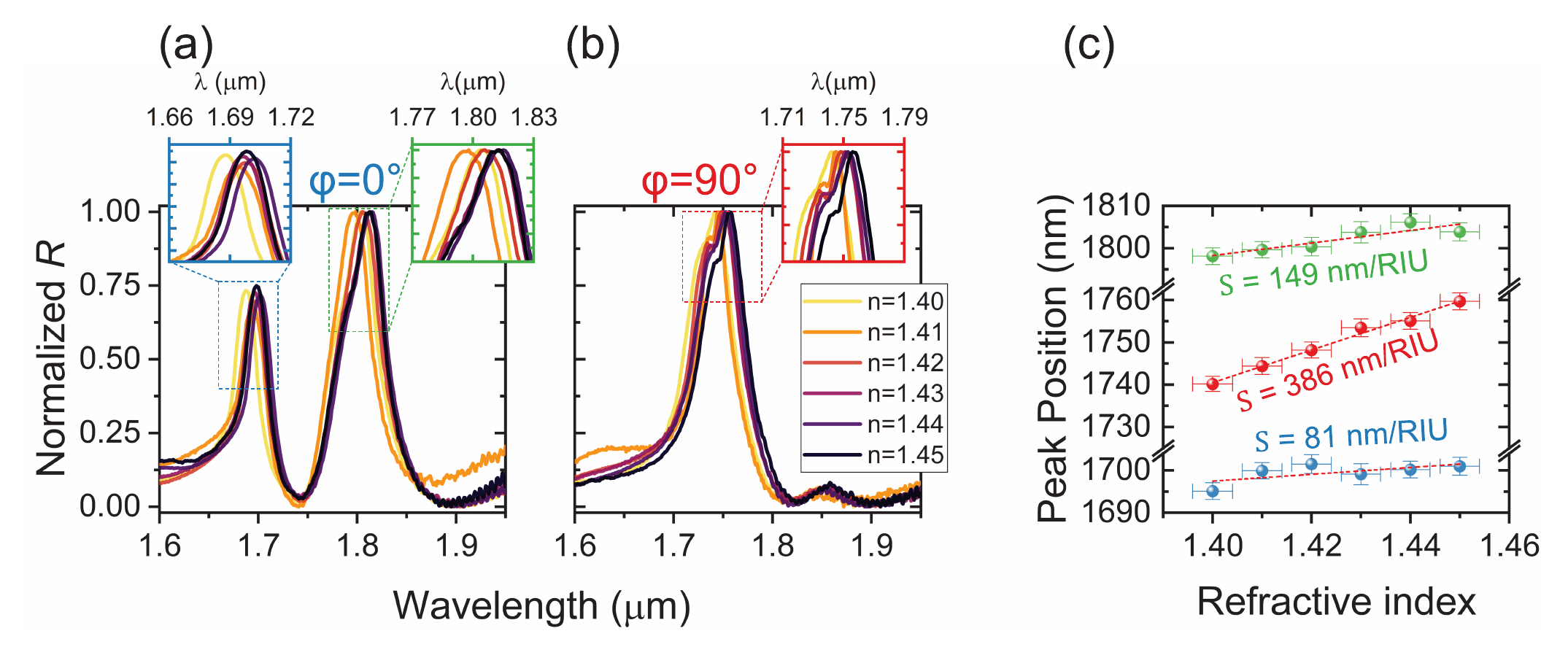}
    \caption{\textbf{RI sensing results for the \ce{Si}/\GeSn{0.9}{0.1} array}. (a) Normalized reflectance spectra for different refractive index solutions from 1.40 to 1.45 at an incident light polarization of $\varphi=\SI{0}{\degree}$. (b) Similar to panel (a) except the polarization angle $\varphi$ is changed to $\SI{90}{\degree}$. The insets in both panels (a) and (b) are Zoom\hyp{}in on the resonances showing the spectral shift in more detail. (c) Spectral shift of the resonances vs. \ac{ri} of the surrounding solutions. Spheres are the experimental data points, extracted by fitting the indicated regions in panels (a) and (b), based on Eq. \ref{eqn:Rfit}. The y\hyp{}axis error bars indicate the uncertainty of the fitting process, with regards to the peak position, evaluated based on Eq. \ref{eqn:Rfit}. The x\hyp{}axis error bar is related to the uncertainty in the \ac{ri}, evaluated based on the manufacturer \ac{ri} quantification. The dashed red traces are linear fits to the data giving a sensitivity $S$ of $\SI{149}{\nm}/\text{RIU}$ and $\SI{81}{\nm}/\text{RIU}$ for the \ac{pir} Fano resonance at a polarization angle $\varphi$ of $\SI{0}{\degree}$ and $\SI{389}{\nm}/\text{RIU}$ when $\varphi=\SI{90}{\degree}$. The $R^2$ of the linear regression is above 0.98.}
    \label{fig:fig5RIsensor}
\end{figure*}
The metasurface is immersed in a \ac{ri}\hyp{}matching oils (\textcolor{blue}{Methods}) to supply different solutions with variable \ac{ri} while the reflectance spectra are recorded at both \textit{s}\hyp{} and \textit{p}\hyp{}polarizations. Due to the strong interference of the \ac{ed} and \ac{md} resonant modes, the Fano resonance mainly depends on the non\hyp{}radiation damping, which makes it sensitive to the changes in the surrounding dielectric environment. Above all, the Fano mode exhibits a distinct resonance shift with respect to the small fluctuation in the \ac{ri} of the surrounding medium $(\Delta n=0.01)$ (\fref{fig:fig5RIsensor}{}). Therefore, this offers an excellent potential for ultrahight resolution required in bio\hyp{}sensing and gas detection. \fref{fig:fig5RIsensor}{a,b} shows the normalized reflectance spectra for the \ce{Si}/\GeSn{0.9}{0.1} \ac{nw} array as a function of variable \ac{ri}, $n$ from $1.40$ to $1.45$ at two distinct polarizations.
According to Eq. \ref{eqn:fom}, the sensitivity $S$ is evaluated through the resonance wavelength shift over the \ac{ri} change unit (RIU), while \ac{fom} takes further consideration of the resonant lineshape \cite{Sherry2005LocalizedNanocubes}. The \ac{fwhm} indicates the ability to confine electromagnetic fields at resonance mode, which set the sensing resolution. To get a consistent estimate for the \ac{fom} value, the \ac{fwhm} are obtained from Eq. \ref{eqn:Rfit} and evaluated for each polarization angle, an \ac{fom} of 5.35 and 6.56 for the \ac{ed} and \ac{md} modes, respectively at $\varphi=\SI{0}{\degree}$. The \ac{fom} is around 8.0 at $\varphi=\SI{90}{\degree}$. The \ac{ed} and \ac{md} $Q$-factors are 57 and 40, respectively at $\varphi=\SI{0}{\degree}$, whereas, at $\varphi=\SI{90}{\degree}$, the $Q$-factor decreases to 30.5 due to the increase in the associated spectral width. The $Q$-factor is defined as $Q=\lambda/\Delta\lambda$, where $\Delta\lambda$ is the spectral width of the corresponding resonant mode. \fref{fig:fig5RIsensor}{c} shows a clear redshift of the Fano resonance peak, as well as the asymmetric \ac{ed} and symmetric \ac{md} resonances with increasing \ac{ri}. The sensitivity $S=\delta\lambda/\delta n\;(nm/\text{RIU})$ of the metasurface is estimated through a linear fit of the peak position as a function of the \ac{ri}. The sensitivity $S$ is evaluated for both \textit{s} and \textit{p} polarizations. At $\varphi=\SI{0}{\degree}$, the sensitivity associated to the \ac{ed} and \ac{md} modes shifts are equal to 81 and $\SI{149}{\nm}/\text{RIU}$, respectively. The small sensitivity is directly related to the asymmetric nature of the Fano resonance, characterized by a negative $q$ parameter $(\sim -2)$ (\textcolor{blue}{Table S1}) , when $\varphi=\SI{0}{\degree}$. Next, at $\varphi=\SI{90}{\degree}$, $S$ increases in average 3\hyp{}fold from the \textit{s}\hyp{} to the \textit{p}\hyp{}polarized incident light. To shed light on the performance of the \ac{ri} sensor, a brief review of the preceding experimental all\hyp{}dielectric \ac{ri} sensors is presented in \textcolor{blue}{Table S2}. Several characteristics distinguish the current work from the current state\hyp{}of\hyp{}the\hyp{}art. The first is the polarization\hyp{}dependent sensitivity $S$ of the \ac{ri} sensor. The second is the \ac{swir} accessible spectral range, which cannot be covered by the reported platforms thus far. Sensing devices operating in this spectral range are sought\hyp{}after for both commercial and defense applications. The third is the achieved high sensitivity of $\SI{386}{\nm}/\text{RIU}$ even though the $Q-$ factor of the resonance is relatively low $(30.5)$, when compared to early reports \cite{yang2014}. The observed increase in sensitivity is directly linked to the change in the asymmetry parameter $q$. Indeed, the more symmetric is the resonance mode, the higher is the sensitivity. In principle, any control of the Fano resonance features is driven by the geometric coupling that is accessible through micro\hyp{}fabrication steps to enhance the sensing sensitivity. Although a metallic metamaterial has generally higher \ac{ri} sensitivity than its all\hyp{}dielectric counterparts, the former has smaller $Q$\hyp{}factor than the latter. This directly affects the \ac{fom} of \ac{ri} sensors. 

\UseRawInputEncoding
In conclusion, we demonstrated the conformal growth of a $\SI{150}{\nm}$\hyp{}thick metastable \GeSn{0.9}{0.1} shell around tapered \ce{Si} nanotips to establish a wafer\hyp{}level, all\hyp{}dielectric metasurface. We also presented a detailed discussion of its polarization\hyp{}dependant specular reflectance. A rigorous theoretical analysis revealed that the interference between the \ac{ed} and \ac{md} modes underlies the observed Fano resonance nature of the \ac{pir}. Additionally, the modulation of the Fano resonance by tuning the polarization is achieved by relating the asymmetry parameter of the Fano lineshape to the coupling strength between the \ac{ed} and \ac{md} moments. To harness these polarization\hyp{}tailored Fano resonances, a refractive index sensor is demonstrated using the fabricated metasurface. The sensitivity $S$ of the proposed device reaches $\SI{386}{\nm}/\text{RIU}$ at a polarization angle of $\SI{90}{\degree}$, whereas at a $\SI{0}{\degree}$ polarization, the \ac{ed}\hyp{} and \ac{md}\hyp{}dependent sensitivity are $\SI{149}{\nm}/\text{RIU}$ and $\SI{81}{\nm}/\text{RIU}$, respectively. It is possible to obtain a higher $Q$\hyp{}factor and further improve the \ac{fom} by tuning only the tapered base dimensions, instead of the whole \ac{nw}. This would simplify the design and fabrication of future metasurfaces. This platform can be easily integrated with microfluidic systems for lab\hyp{}on\hyp{}a\hyp{}chip applications \cite{Peyskens2016,Chen2018}. Tunable Fano resonances in the \ac{swir} range using all\hyp{}dielectric metasurfaces pave the way to numerous applications ranging from nonlinear optics and sensing to the realization of new types of optical modulation and low\hyp{}loss slow\hyp{}light devices.\\
\medskip

\textbf{METHODS} 
\par \textbf{Array Fabrication:} Using a state\hyp{}of\hyp{}the\hyp{}art pilot line for $\SI{130}{\nm}$ \SiGe{}{} BiCMOS technology, \ce{Si} \ac{nw} with diameters of 20\hyp{}$\SI{150}{\nm}$ (top\hyp{}bottom) were fabricated from a $\SI{300}{\mm}$ \ce{Si}(100) wafer by a multi\hyp{}step procedure, including a anisotropic \ce{Ar}:\ce{BCl3}:\ce{Cl2}\hyp{}based reactive ion etching (RIE) process \cite{skibitzki2017}. The \ce{Si} \ac{nw}s sample was then cleaned in an \ce{HF}\hyp{}based solution prior to loading in the \ac{cvd} reactor. The \GeSn{}{} shell was grown at $\SI{300}{\degreeCelsius}$ using ultra\hyp{}pure \ce{H2} carrier gas, and 10 \% monogermane (\ce{GeH4}) and tin\hyp{}tetrachloride (\ce{SnCl4}) precursors. A constant precursor supply with a \ce{GeH4}/\ce{SnCl4} ratio of $\sim$ 1700 was provided during the \GeSn{}{} growth, hence with the same parameters as in the reference \GeSn{0.90}{0.10} thin film layers grown on a \ce{Ge} on \ce{Si} substrate \cite{Assali2019a}.
\par \textbf{Optical Measurements:} Polarization\hyp{}resolved reflectance was acquired from $\SI{1}{\um}$ to $\SI{2.5}{\um}$ using a focused beam with a spectrophotometer. The wavelength was scanned at a step of $\SI{1}{\nm}$. The incident angle was varied between $\SI{6}{\degree}$ and $\SI{52}{\degree}$ with a step of $\SI{1}{\degree}$ and the incident light polarization was changed between \textit{s}\hyp{} and \textit{p}\hyp{}states, as detailed by the schematic in \fref{fig:fig2eit}{a}. A dispersion map of the metasurface was also measured (\textcolor{blue}{Fig. S2}). The spot size was approximately $\SI{2.5}{\milli\meter}$ $\times$ $\SI{2.5}{\milli\meter}$, which covered approximately $\SI{20}{\million}$ \ac{nw}s. The wire grid polarizer was rotated from \textit{s}\hyp{} to \textit{p}\hyp{}polarized state with a $\SI{10}{\degree}$ step. The \ac{nw} metasurface was immersed in the refractive\hyp{}index\hyp{}matching oils (Cargille Labs) for all the sensing measurements. The samples were washed thoroughly after each measurement with isopropyl alcohol, dried with nitrogen gas, and followed by placing them in a vacuumed desiccator for \SI{30}{\minute} to ensure that no residual oil was left after each measurement.

\par \textbf{Simulations:} The \ac{fdtd} calculations are performed using Ansys\hyp{}Lumerical$^\copyright$ software. During the calculations, an electromagnetic pulse in the wavelength range from $1.1$ to $\SI{2.5}{\um}$ is launched into a box containing the target core/shell \ac{nw} to simulate a propagating plane wave interacting with the \ac{nw}. The polarization type is defined with respect to the electromagnetic field orientation (\textit{p}\hyp{}type when the electric field is parallel to the surface,\ $\varphi=\SI{0}{\degree}$, and \textit{s}\hyp{}type when the magnetic field is parallel to the surface,\ $\varphi=\SI{90}{\degree}$). The \ce{Si}/\GeSn{}{} array and its surrounding space are divided into $\SI{5}{\nm}$ meshes. The refractive index of the medium in the top and side regions is that of air $(n=1)$ and that in the bottom is set according to the dielectric functions of \ce{Si} substrate. The refractive index of the silicon core is approximated to that of a \ce{Si} substrate. A reference \GeSn{0.9}{0.1} thin film was epitaxially grown to quantify both its real and imaginary dielectric function with spectroscopic ellipsometry \cite{Attiaoui2021}. A detailed structural characterization of the reference layer is shown in \textcolor{blue}{Fig. S11}. The unit cell core/shell \ac{nw} is modeled as two concentric frustums, as clearly highlighted in the \ac{stem} images (\textcolor{blue}{Fig. S3}). Its dimensions are set to be the same as the average size measured from the \ac{sem} images. The different excitation polarizations are considered by setting the electric field in the substrate plane either parallel or perpendicular to the \ac{nw} length axis, as well as being vertical to the substrate. Power monitors, positioned $\SI{100}{\nm}$ below the air\hyp{}\ce{Si} interface and $\SI{500}{\nm}$ above the \ac{nw}, are used to determine the transmission and reflection at each wavelength, respectively. \ac{pml} boundary conditions are used in the vertical direction to prevent nonphysical scattering at the edge of the simulation box. When \ac{aoi}$=\SI{0}{\degree}$, periodic boundary conditions are used in both in\hyp{}plane dimensions to simulate an infinite periodic array, whereas Bloch boundary condition are considered when $\text{AOI}>\SI{0}{\degree}$. The field distributions are obtained according to the electric and magnetic field distributions at different resonance modes. They are calculated on the cross\hyp{}sectional plane that passes through the \ac{nw} z\hyp{}axis and perpendicular to the substrate (along x\hyp{}y plane).

\par \textbf{Scattering Calculation:} The scattering\hyp{}cross section simulations are performed using \ac{fdtd} software (Ansys\hyp{}Lumerical$^{\copyright}$) running on a $\SI{5}{\giga\hertz}$ workstation. The mesh size around the core/shell \ac{nw} in the simulations is $5\times 5\times \SI{5}{\nm\cubed}$. The \ac{tfsf} plane wave is used as the excitation source, which leads to accurate evaluation of the scattered field outside the metasurface and the corresponding scattering cross section. \ac{pml} boundary conditions are used to simulate an individual structure placed in an infinite space. The surrounding index in the simulations is vacuum with a refractive index $n = 1$.

\par \textbf{Fitting Models:} Two fitting models are considered in this work. First, a lineshape fitting approach is employed to extract the $Q$\hyp{}factor, \ac{fwhm} and mode resonance from the measured polarization\hyp{}dependent reflectance ratio. The total reflectance $R(\omega)$ is expressed in the following form 
\begin{myequation}{1.1}
    \begin{aligned}
    R(\omega) &= \frac{a^2}{\vertarrowbox{\left(\frac{\omega^2-\omega_1^2}{2\gamma_1\omega_1}\right)^2+1}{$R_1(\omega)$}} \cross \frac{\left(\frac{\omega^2-\omega_2^2}{2\gamma_2\omega_2}+q\right)^2+b}{\vertarrowbox{\left(\frac{\omega^2-\omega_2^2}{2\gamma_2\omega_2}\right)^2+1}{$R_2(\omega)$}}
    \label{eqn:Rfit}
    \end{aligned}
\end{myequation}
The multipliers represent an interference between the radiation continuum and the \ac{ed} mode throughout a symmetric Lorentzian lineshape $R_1(\omega)$ and the radiation continuum coupled to the \ac{md} mode through an asymmetric Fano\hyp{}like lineshape $R_2(\omega)$. The asymmetric lineshape $R_2(\omega)$ describes the destructive or constructive interference due to the coherent coupling between the \ac{ed} and \ac{md} modes. $\omega_1,\gamma_1$ and $\omega_2,\gamma_2$ are the resonant frequency and an approximation of the resonance spectral width of the \ac{ed} and \ac{md} resonances, respectively. $q$ is the Fano asymmetry parameter, $b$ is the damping parameter originating from intrinsic losses, and $a$ is the maximal amplitude of the resonance.
Second, the \ac{cho} model is employed to quantify the coupling strength between the two modes. The fit is performed over the wavelength range from $\SI{1.6}{\um}$ to $\SI{1.9}{\um}$, by solving the following coupled second\hyp{}order differential equations, 
\begin{myequation}{1.1}
    \begin{aligned}
    \ddot{x}_1+\gamma_1\dot{x}_1+\omega_1^2x_1+\tilde{\kappa}'\dot{x}_2 &= \alpha E_0e^{i\omega t}\abs{\cos{\varphi}},\\
    \ddot{x}_2+\gamma_2\dot{x}_2+\omega_2^2x_2-\tilde{\kappa}'\dot{x}_1 &= 0,
    \label{eqn:CHO}
    \end{aligned}
\end{myequation}

The first oscillator models the \ac{ed} (radiative) mode with a frequency $\omega_1$ and a damping $\gamma_1$ representing radiative losses. The second oscillator emulates the \ac{md} (nonradiative) with a resonant frequency $\omega_2$ and a smaller damping $\gamma_2$. The two oscillators coupling coefficient is $\tilde{\kappa}'$ and $\alpha$ is a parameter indicating the coupling strength of the radiative \ac{ed} mode with the incident electromagnetic field $E_0e^{i\omega t}$. A more elaborate discussion in relation to the effect of the coupling strength on the reflectance ratio, is detailed in \textcolor{blue}{Fig. S8}.

\bigskip \medskip
\paperSection{ACKNOWLEDGEMENTS}
\indent The authors thank J. Bouchard for the technical support with the \ac{cvd} system, Brett Carnio with help with the Ansys\hyp{}Lumerical$^{\copyright}$ simulations, and C. Lemieux-Leduc for help with the Blender 3D schematic image in \fref{fig:fig1meta}{a}. O.M. acknowledges support from NSERC Canada (Discovery, SPG, and CRD Grants), Canada Research Chairs, Canada Foundation for Innovation, Mitacs, PRIMA Qu\'ebec, and Defense Canada (Innovation for Defense Excellence and Security, IDEaS).\\

\paperSection{AUTHORS INFORMATION}
Corresponding Authors:\\
\textcolor{blue}{$^{*}$}\href{mailto:anis.attiaoui@polymtl.ca}{anis.attiaoui@polymtl.ca}\\\textcolor{blue}{$^{\dagger}$} \href{mailto:oussama.moutanabbir@polymtl.ca}{oussama.moutanabbir@polymtl.ca}\\
Notes:\\
The authors declare no competing financial interest.

\bigskip

\let\oldaddcontentsline\addcontentsline
\renewcommand{\addcontentsline}[3]{}

\bibliography{main.bib} 
\bibliographystyle{apsrev4-2} 



\ifthenelse{\equal{\includeSI}{1}}{
	\input{customCommands_SI.tex}
    \renewcommand{\tocname}{Table of contents}
     \setcounter{figure}{0}\setcounter{table}{0}
     \renewcommand{\theHtable}{Supplement.\thetable}
    \renewcommand{\theHfigure}{Supplement.\thefigure}
     \let\addcontentsline\oldaddcontentsline
     \blankpage
	\include{main_SI_body.tex}
}{}

\end{document}